# Evolution of structure of SiO$_2$ nanoparticles upon cooling from the melt

Nguyen Thi Xuan Huynh*[1], Vo Van Hoang*[2] and Hoang Zung[3]

Address: [1]Department of Physics, Quy Nhon University, Vietnam, 170 An Duong Vuong Street, Quy Nhon City, Binh Dinh Province, Vietnam, [2]Department of Physics, Institute of Technology of HochiMinh City-Vietnam, 268 Ly Thuong Kiet Street, District 10, HochiMinh City-Vietnam and [3]Department of Science and Technology, National University of HochiMinh City, Vietnam

Email: Nguyen Thi Xuan Huynh* - huynhqn@gmail.com; Vo Van Hoang* - vvhoang2002@yahoo.com; Hoang Zung - hdung@vnuhcm.edu.vn

* Corresponding authors

## Abstract

Evolution of structure of spherical SiO$_2$ nanoparticles upon cooling from the melt has been investigated via molecular-dynamics (MD) simulations under non-periodic boundary conditions (NPBC). We use the pair interatomic potentials which have weak Coulomb interaction and Morse type short-range interaction. The change in structure of SiO$_2$ nanoparticles upon cooling process has been studied through the partial radial distribution functions (PRDFs), coordination number and bond-angle distributions at different temperatures. The core and surface structures of nanoparticles have been studied in details. Our results show significant temperature dependence of structure of nanoparticles. Moreover, temperature dependence of concentration of structural defects in nanoparticles upon cooling from the melt toward glassy state has been found and discussed.

**PACS Codes**: 61.43.Bn; 78.55.Qr; 78.67.Bf

## Introduction

Silica nanoparticles have potential applications in many fields including ceramics, chromatography, catalysis and chemical mechanical polishing [1]. In recent years, SiO$_2$ nanoparticles have been investigated by means of experimental techniques such as NMR (nuclear magnetic resonance), SAXS (small angle X-ray scattering) [2], light absorption [3], FTIR (Fourier transform infrared) spectra and photoluminescence [4-7], etc. According to the diffraction data [8], liquid and amorphous silica have $Z_{Si-O}$ = 4 and $Z_{O-Si}$ = 2, i.e. the main structural element of the network is a slightly distorted SiO$_4$ tetrahedron and the adjacent tetrahedra are linked to each other through the shared vertices. This means that Si atoms having 1, 2, 3 or 5-fold coordinations and O atoms have 1 or 3-fold coordinations can be considered as structural defects in $SiO_2$ nanoparticles, which are similar to those observed in the bulk [9,10]. Among them, the most serious defects are structural units with $Z_{si-O}$ = 2, $Z_{si-O}$ = 3 and $Z_{si-O}$ = 1. It was found that the structural defects can play an important role in their structure and properties including photoluminescence [6], catalysis and micro-electronics [11]. In particular, the structural defects lead to the narrowing of band gap and the formation of localized states within band tail, which were experimentally measured from light-absorption study of silica nanoparticles [3]. However, very few theoretical researches have been done for $SiO_2$ nanoparticles despite the observation of interesting properties [11-16]. Therefore, it motivates us to carry out the research on liquid and amorphous $SiO_2$ nanoparticles by MD simulation via the microstructural analysis. Namely, we investigate the structural evolution of $SiO_2$ nanoparticles upon cooling from the melt toward glassy state (or amorphous one).

**Calculation**

In the present study, we use interatomic potentials including weak Coulomb interactions and the Morse type potential for short-range interactions as given bellow:

$$U_{ij}(r) = \frac{q_i q_j}{r} + D_0 \{\exp[\gamma(1 - \frac{r}{R_0})] - 2\exp[\frac{1}{2}\gamma(1 - \frac{r}{R_0})]\}$$

Where $q_i$ and $q_j$ represent the charges of ions $i$ and $j$, for Si atom $q_{Si}$ = 1.30$e$ and for O atom $q_0$ = -0.65$e$ ($e$ is the elementary charge unit); $r$ denotes the interatomic distance between atoms $i$ and $j$; the other parameters of the Morse potentials can be found in Refs. [17-19]. MD simulations were done in a spherical $SiO_2$ particle with three different sizes of 2 nm, 4 nm and 6 nm corresponding the real density of 2.20 g/cm$^3$ for amorphous $SiO_2$ which have corresponding numbers of atoms of 276 (92 silicon atoms and 184 oxygen ones), 2214 (738 silicon atoms and 1476 oxygen ones) and 7479 (2493 silicon atoms and 4986 oxygen ones), respectively. We used the Verlet algorithm with the MD time step of 1.60 fs. Each model contained the number of Si and O atoms in accordance to the $SiO_2$ stoichiometry. Firstly, N atoms are randomly placed in a sphere of fixed radius and the NPBC model has been relaxed for $5 \times 10^4$ MD steps at 7000 K in order to get a good equilibrated liquid model. We use the simple non-slip with non-elastic reflection behavior boundary (i.e. if during the relaxation atoms move out of the spherical boundary they have to be placed back to the surface and it is called NPBC). Then, the temperature of the system was decreased linearly in time as $T = T_0 - \alpha t$ where $\alpha = 4.2956 \times 10^{13}$K/s is the cooling rate, $T_0$ is the initial temperature of 7000 K and $t$ is the cooling time. This cooling process was continued until the temperature of the system was equal to 350 K. It is essential to notice that melting and evaporation points for bulk $SiO_2$ are $T_m$ = 1923K and $T_{boil}$ = 2503K, respectively [1]. This means that the initial configuration at $T$ = 7000K corresponds to the superheated liquid and that only due to using of NPBC it is possible to reach such a condition for $SiO_2$ nanoparticles otherwise the evaporation occurs. Note that using potentials have been successfully used in MD simulations of both structure and thermodynamic properties of silica [17-19], and in the investigation of the structure changes in cristobalite and silica glass at high temperatures [18]. These potentials reproduced well the melting temperature of cristobalite and the glass phase transition temperature of silica glass and calculated data were more accurate than those observed in other simulation works in which the traditional interatomic potentials with more strong electrostatic interaction have been used such as BKS or TTAM potentials [20-22]. Although the potentials described above were proposed for the bulk $SiO_2$, they also described well structural features and surface energy of amorphous $SiO_2$ nanoparticles compared with those observed for amorphous $SiO_2$ nanoclusters by using BKS ones [15]. Melting point of nanoparticles is size dependent in that it strongly reduces with decreasing nanoparticle size [23], and it is out of scope of the present work. In order to investigate the evolution of structure of $SiO_2$ nanoparticles upon cooling from the melt, we saved a number of configurations at finite temperatures (i.e. at temperatures ranged from 7000 K to 350 K with the temperature interval of 700 K). and then we relaxed them for $5 \times 10^4$ MD steps before calculating static properties. In order to calculate the coordination number and bond-angle distributions in $SiO_2$ nanoparticles, we adopted the fixed values $R_{Si-Si}$ = 3.30 Å, $R_{Si-O}$ = 2.10 Å and $R_{O-O}$ = 3.00 Å. Here $R$ corresponds to the position of the minimum after the first peak in PRDFs for the amorphous state at a real density of 2.20 g/cm$^3$. In order to improve statistics, the results have been averaged over four and three independent runs for nanoparticles with the size of 2 nm and 4 nm, respectively. The single run was done for the size of 6 nm due to large number of atoms in the model. In addition, since surface structure plays an important role in the structure and properties of nanoparticles we also focus attention to the surface of $SiO_2$ nanoparticles. Therefore, we need a criterion to decide which atoms belong to the surface and which ones belong to the core of nanoparticles. There is no common principle for such choice of surface or core of the amorphous substances. The definition of thickness of the surface in [13] is somewhat arbitrary: all atoms that were within 5 Å of the hull just touched the exterior of the droplet, and were considered to belong to the surface, atoms that had the distance between 5 Å and 8 Å from the hull belong to the transition zone and the remaining atoms belong to the interior. In contrast, no definition of surface was clearly presented for the amorphous $Al_2O_3$ thin film; they used the top 1 Å or 3 Å layer of the amorphous thin film for surface structural studies [24]. From structural point of view, it can be considered that atoms belong to the surface if they could not have full coordination for all atomic pairs in principle and in contrast, atoms belong to the core if they could have full coordination for all atomic pairs in principle like those located in the bulk. Therefore, in the present work atoms located in the outer shell of $SiO_2$ spherical nanoparticle with thickness of 3.30 Å (i.e. the largest radius of the coordination spheres found in the system) belong to the surface of the nanoparticles and the remaining atoms belong to the core.

**Results and discussion**

The first quantity we would like to discuss here is the PRDFs, $g_{ij}(r)$, for Si-Si, Si-O and O-O pairs at three different temperatures of 7000 K, 3500 K and 350 K. Fig. 1 shows that $g_{ij}(r)$ of $SiO_2$ nan-oparticles obtained at $T$ = 350K is typical for a glassy state (or amorphous one) of the system like those found for the bulk amorphous $SiO_2$ obtained by using different interatomic potentials or by experiment [9,10,25] since we use very high cooling rate of $\alpha$ = 4.2956 × $10^{13}$K/s. Furthermore, Fig. 1 & Table 1 show that the peaks in PRDFs strongly depend on the temperature, i.e. they become more pronounced upon cooling in that the changes in first peaks and minima are the most remarkable. More detailed changes in structure can be found via the coordination number distributions (Tables 1 &2). One can see that when the temperature decreases, mean coordination number for all pairs, $Z_{ij}$, increases (Fig. 2). In particular, $Z_{si-O}$ shifts from 3 to 4 (Tables 1 &2). This means that tetrahedral network structure of $SiO_2$ nanoparticle is forming upon cooling and at the temperature of 350 K (amorphous sate), about 93.13% Si atoms are surrounded by 4 oxygen atoms. These results agree with those obtained via computer simulations for liquid and amorphous $SiO_2$ by using BKS potentials for the system at lower temperatures (about 99%) [9] or by using the potentials which have weak Coulomb interaction and Morse type short-range interaction for the bulk (about 97.3%) [10]. Indeed, in our models at low temperature of 350 K it is also found that the mean coordination numbers $Z_{Si-O} \approx 4$ and $Z_{O-Si} \approx 2$.

As mentioned in the introduction, the structural defects in $SiO_2$ nanoparticles can play an important role in structure and properties of silica nanoparticles [6,11]. Therefore, it is interesting to study the evolution of coordination number distributions for Si-O and O-Si pairs upon cooling from the melt. Table 2 shows that the number of defects with $Z_{si-O}$ = 2 and $Z_{O-Si}$ = 1 decreases with decreasing temperature. In contrast, the number of Si atoms having $Z_{si-O}$ = 3 (3-fold coordination) decreases with decreasing temperature only after about 4200 K. The defects having 1 or 5-fold coordinations for Si atoms and 3-fold coordinations for O atoms also decrease with decreasing temperature (not shown due to small fraction). Moreover, our calculations show that the probability for the occurrence of structural defects is described well by an Arrhenius law $P_{ij} = A_{ij} \exp(-E_{ij}/T)$, like those were discussed for the bulk in [9,10]. We found that $P_{ij}$ is described well by such law in temperature ranges such as: 2100K–4200K for $SiO_3$ ($Z_{Si-O}$ = 3), 2800K–7000K for $SiO_2$ ($Z_{si-O}$ = 2) and 2100K–7000K for $Z_{O-Si}$ = 1 (Fig. 3). Here $P_{ij}$ denotes the probability that an $i$-type ion exactly has $Z$ nearest neighbors of type $j$, $E_{ij}$ denotes activation energies. By extrapolation, we found prefactors $A_{ij}$ and $E_{ij}$ for 2 nm, 4 nm and 6 nm particles, which are shown in Table 3.

**Table 1: Structural characteristics of 4 nm $SiO_2$ nanoparticle upon cooling from the melt.**

| $T$ (K) | $r_{ij}$ | | | $g_{ij}$ | | | $Z_{ij}$ | | | |
|---|---|---|---|---|---|---|---|---|---|---|
| | Si-Si | Si-O | O-O | Si-Si | Si-O | O-O | Si-Si | Si-O | O-Si | O-O |
| 7000 | 2.98 | 1.51 | 2.61 | 2.90 | 9.29 | 2.79 | 2.27 | 3.00 | 1.56 | 3.97 |
| 3500 | 3.02 | 1.52 | 2.58 | 4.52 | 14.93 | 4.08 | 2.73 | 3.59 | 1.80 | 4.83 |
| 350 | 3.02 | 1.52 | 2.54 | 8.86 | 32.22 | 9.30 | 3.76 | 3.93 | 1.96 | 5.96 |
| Exp. [8] | 3.12 | 1.62 | 2.65 | | | | 4.00 | 4.00 | 2.00 | 6.00 |

Here $r_{ij}$ – The position of the first peaks in PRDFs; $g_{ij}$ – The height of the first peaks in PRDFs; $Z_{ij}$ – The mean coordination number.

Table 2: Percentage of Si and O atoms with corresponding coordination numbers for $SiO_2$ nanoparticle with the size of 4 nm upon cooling from the melt to the amorphous state.

| T (K) | $Z_{Si-O}$ | | | $Z_{O-Si}$ | |
|---|---|---|---|---|---|
|  | 2 | 3 | 4 | 1 | 2 |
| 7000 | 21.68 | 53.25 | 22.13 | 46.25 | 51.33 |
| 6300 | 16.58 | 55.51 | 25.93 | 43.52 | 53.86 |
| 5600 | 13.47 | 54.27 | 30.41 | 41.13 | 56.30 |
| 4900 | 7.90 | 53.93 | 36.50 | 35.46 | 62.12 |
| 4200 | 4.74 | 50.77 | 42.82 | 30.90 | 67.10 |
| 3500 | 1.99 | 37.40 | 59.44 | 22.09 | 75.71 |
| 2800 | 0.72 | 23.98 | 74.75 | 13.87 | 84.70 |
| 2100 | 0.72 | 11.20 | 87.85 | 6.98 | 92.16 |
| 1400 | 0.72 | 6.82 | 92.32 | 4.97 | 94.08 |
| 350 | 0.72 | 6.01 | 93.13 | 4.95 | 93.72 |

The different values for $A_{ij}$ and $E_{ij}$ for $SiO_2$ nanoparticles with three different sizes indicate the surface effects on the concentration and temperature dependence of concentration of structural defects in the systems since the ratio $S/V$ increases with decreasing nanoparticle size ($S$ and $V$ are the surface and volume of nanoparticle, respectively). These values are smaller than those observed in the bulk [10]. Note that similar results have been found for the evolution of structure of $SiO_2$ nanoparticles with three different sizes of 2, 4 and 6 nm. However, for simplicity we show the data in Figures or Tables mainly for 4 nm nanoparticles. In addition, according to our recent calculations point defects (i.e. structural units with $Z_{Si-O} \neq 4$, and $Z_{O-Si} \neq 2$) mostly concentrate in the surface shells of amorphous $SiO_2$ nanoparticles rather than in the core due to breaking bonds at the surface and fraction of structural defects increases with decreasing nanoparticle size because of enhancement of $S/V$ ratio [15]. One more type of structural defects in amorphous nanoparticles is vacancy, i.e. due to specific disordered structure there is a significant amount of large pores which can exchange the position with the nearest neighbor atoms and they can act as vacancies in the diffusion processes like those found and discussed for amorphous $Al_2O_3$ [26]. This means that surface of $SiO_2$ amorphous nanoparticles contains large amount of structural point defects such as vacancies or those with $Z_{si-O} < 4$ and $Z_{O-Si} = 1$ (see Fig. 3), while the core has overcoordinated structural point defects (i.e. $Z_{si-O} = 5$ and $Z_{O-Si} = 3$, for example) if any. Probably, due to small dimension amorphous nanoparticles can have only such structural defects unlike those observed in crystalline nanosized substances [27]. Structural defects at the surfaces of amorphous nanoparticles might enhance diffusion of atomic species like those observed and discussed for silica nanoclusters [12,13]. It may be an origin of different surface properties of amorphous nanoparticles including catalysis, adsorption, optical properties etc. like those observed for nanocrystalline $TiO_2$ (see [27] and references therein). Indeed, strong red photoluminescence of amorphous $SiO_2$ nanoparticles has been attributed to the defects at their inner surfaces [6] and intrinsic point defects are the origin of optical band gap narrowing in fumed silica nanoparticles [28].

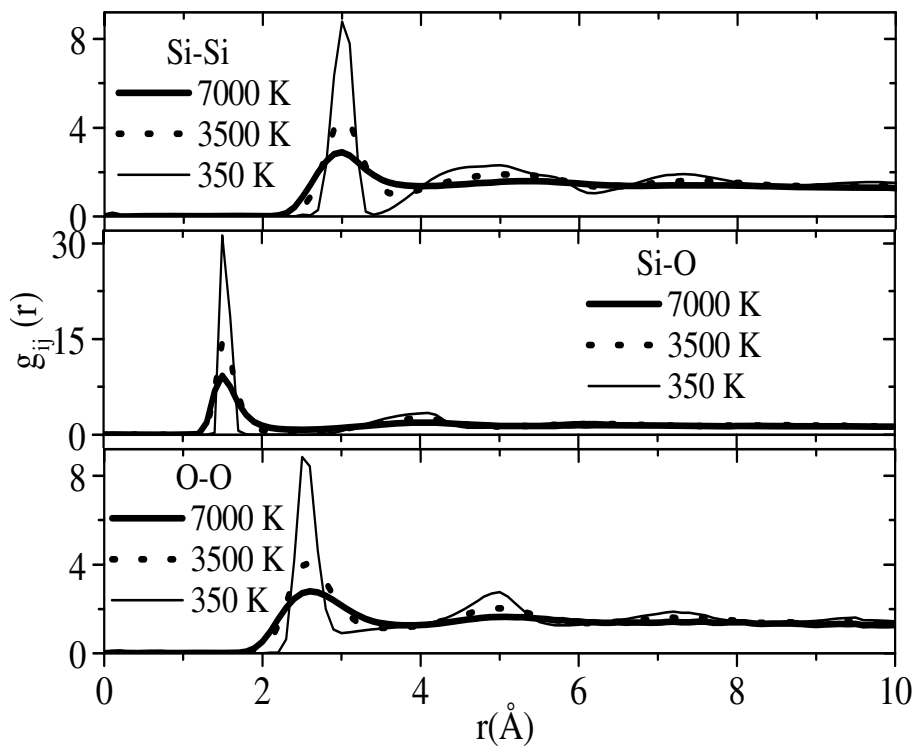

**Figure 1**
Partial radial distribution functions of 4 nm $SiO_2$ nanoparticle upon cooling from the melt.

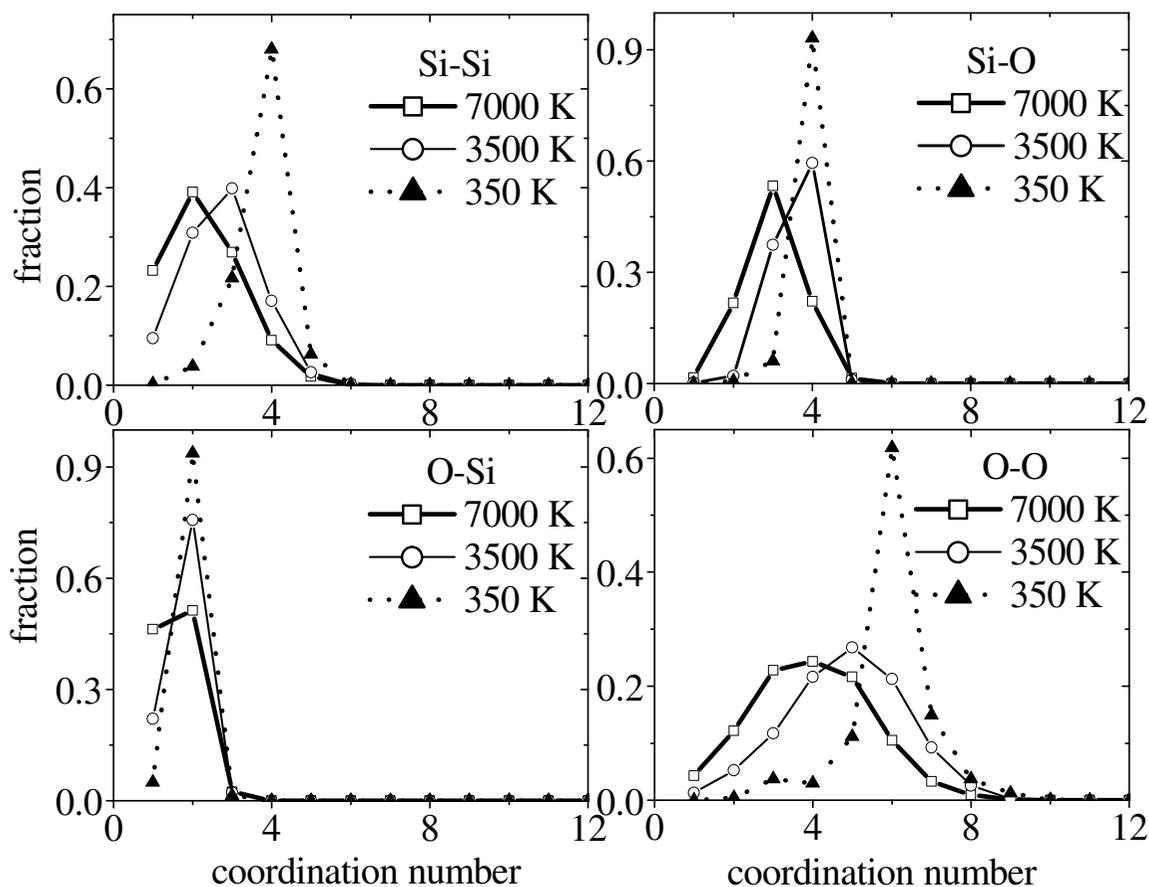

**Figure 2**
Coordination number distributions of 4 nm $SiO_2$ nanoparticle upon cooling from the melt.

**Table 3: Parameters fitted with the Arrhenius law of the probability for the occurrence of structural defects upon cooling from the melt.**

| Size | $Z_{Si-O} = 2$ | | $Z_{Si-O} = 3$ | | $Z_{O-Si} = 1$ | |
|---|---|---|---|---|---|---|
| | $A_{ij}$ | $E_{ij}$ (K) | $A_{ij}$ | $E_{ij}$ (K) | $A_{ij}$ | $E_{ij}$ (K) |
| 2 nm | 1.884 | 13052.693 | 1.781 | 4870.674 | 0.954 | 4033.876 |
| 4 nm | 2.223 | 16215.199 | 2.435 | 6515.887 | 1.162 | 5919.718 |
| 6 nm | 3.273 | 18964.670 | 2.782 | 7391.125 | 1.303 | 6814.464 |
| Data in [9] | | | 58.600 | 31100.000 | 8.900 | 24760.000 |

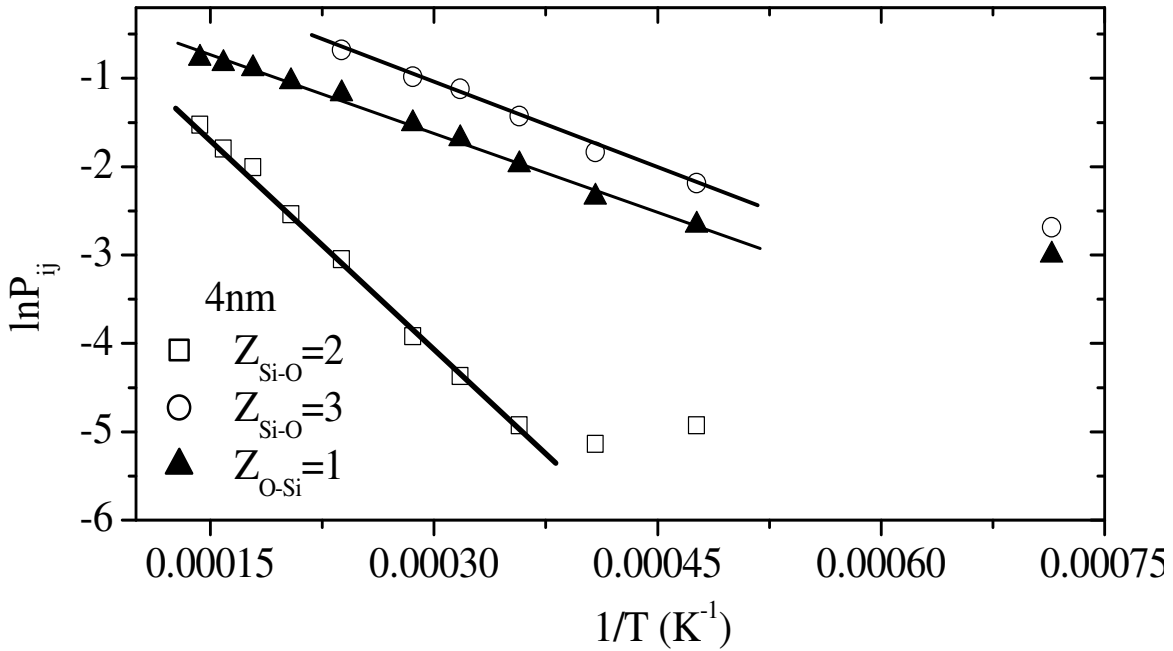

**Figure 3**
Temperature dependence of the probability for the occurrence of structural defects of $SiO_2$ nanoparticle with the size of 4 nm.

More details about the local structure can be inferred from the bond-angle distributions. We present only two most important angles (i.e. Si-O-Si and O-Si-O angles) (Fig. 4). Here, the former describes the connectivity between structural units $SiO_n$ in the system and the latter describes the order inside them. From Fig. 4, we can see that upon cooling from the melt mean bond-angles increase toward larger value (105°–109° for O-Si-O angle and 127°–139° for Si-O-Si angle), close to the O-Si-O angle of an ideal tetrahedral network structure (109.47°). Moreover upon cooling from the melt to amorphous state, our results are close to the experimental values for amorphous silica at the ambient pressure, for which the O-Si-O angle is 109.5° and the Si-O-Si angle is 144° [8]. These changes are in good accordance with the changes in coordination number distributions.

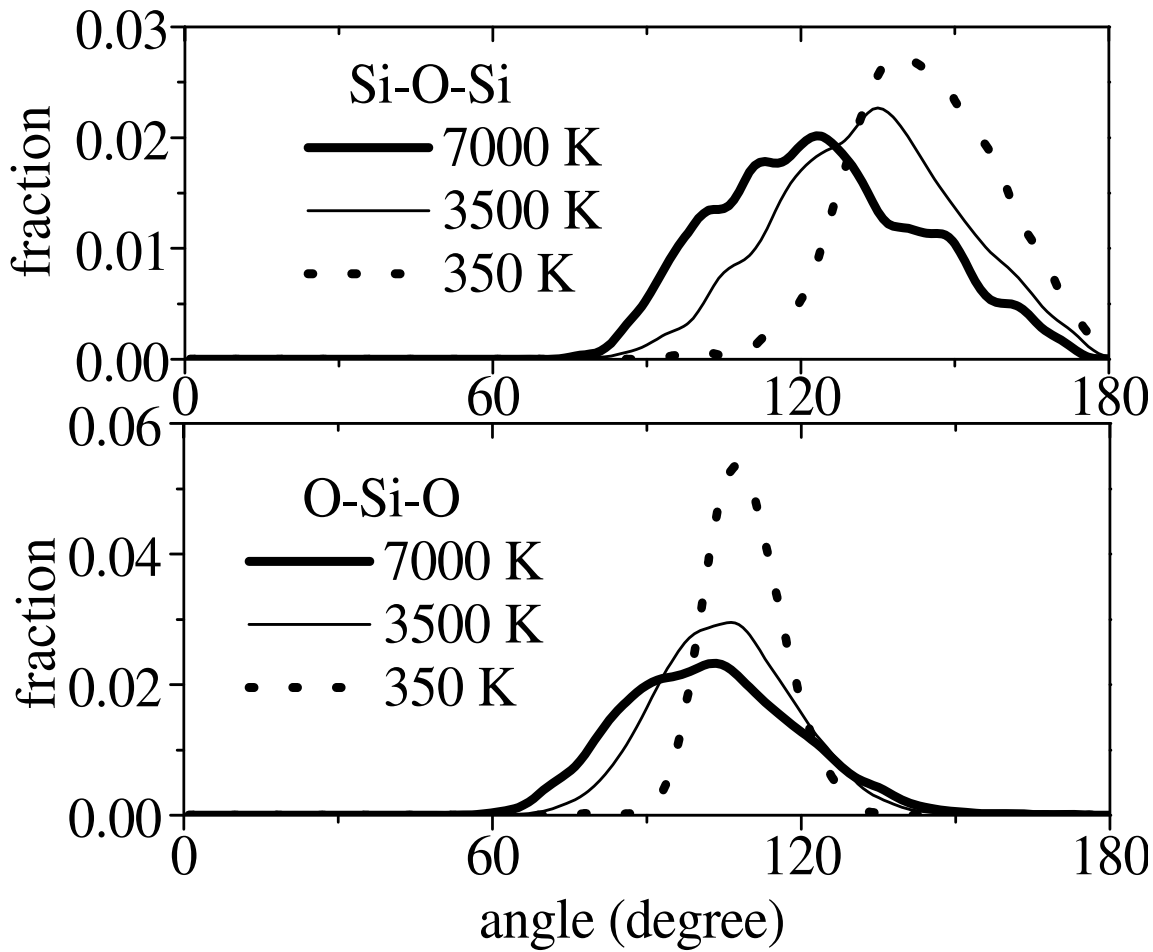

**Figure 4**
Bond-angle distributions of $SiO_2$ nanoparticle with the size of 4 nm upon cooling from the melt.

In order to reveal more clearly about structure of $SiO_2$ nanoparticles, we also show structure of surface and core of nanoparticles upon cooling from the melt. Calculations show that upon cooling from the melt toward glassy state, total number of atoms in the core of $SiO_2$ nanoparticles increases while it decreases in the surface shell. This phenomenon reflects the corresponding change not only in the mass density but also in concentration of defects in two parts of nanoparticles (Fig. 5). It may also cause the change in the stoichiometry (Fig. 6). As is shown in Fig. 6, the discrepancy of stoichiometry in the core and in the surface shell of $SiO_2$ nanoparticles is very clearly, i.e. while surface has a positive deviation from the $SiO_2$ stoichiometry, the core has a negative one over wide range of high temperature region (i.e. in the liquid state). The situation changes slightly into an opposite direction at lower temperatures. This means that at high temperatures oxygen-dificiency defects mainly exist in the core of nanoparticles rather in the surface shell due to the negative deviation of stoichiometry from the standard one (i.e. from $SiO_2$ stoichiometry). In contrast, one can find the existence of 'oxygen-excess defects' in the surface shell due to the positive deviation of stoichiometry from the standard one (see Fig. 6).

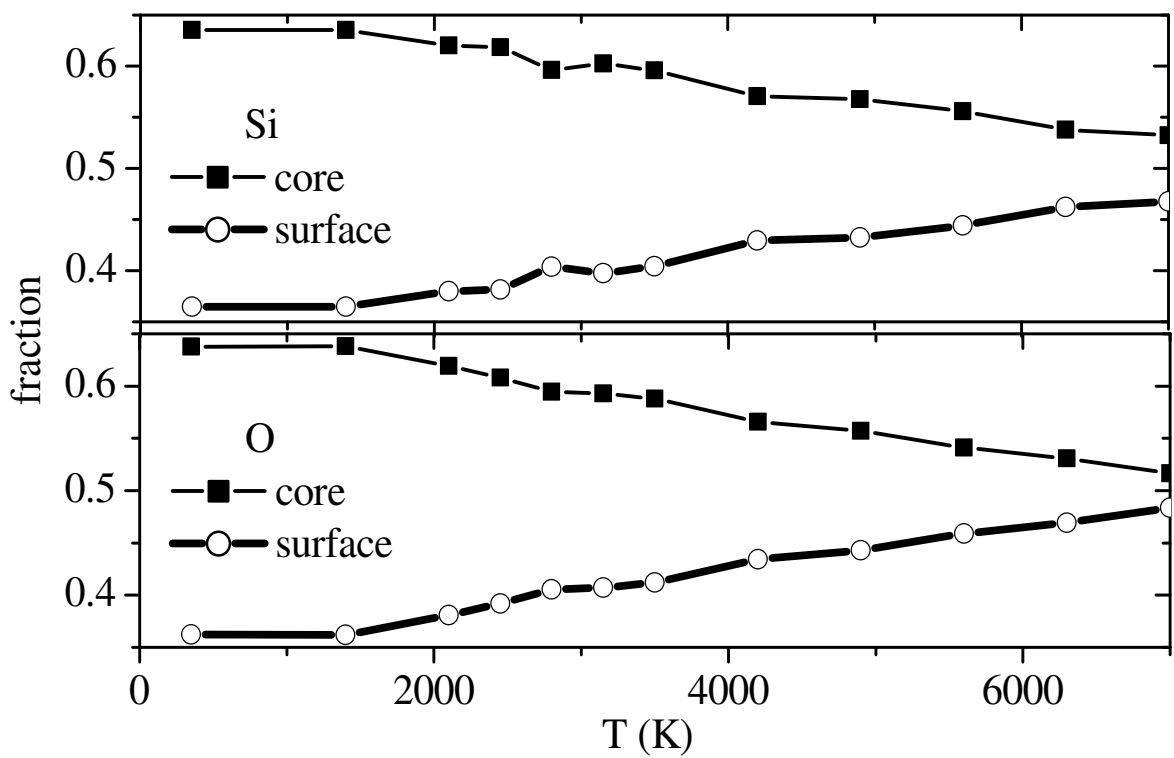

**Figure 5**
Temperature dependence of fraction of Si and O atoms in the core and surface shell of $SiO_2$ nanoparticle with the size of 4 nm.

It is interesting to discuss about the influence of the choice of interatomic potentials and boundary conditions on the obtained results. First, basing on the results obtained for the bulk $SiO_2$ by using different interatomic potentials in literature one can infer that static properties of models are less sensitive dependent on the interatomic potentials used in simulation compared those of the dynamic or thermodynamic ones [9,10,29]. Similar results can be suggested for $SiO_2$ nanoparticles, however, it is also of interest to expand the research in this direction. Quite other situation for the influence of the boundary on the properties of nanoparticles may be occurred. Indeed, according to the model developed in [30], the size dependence of glass transition temperature of nanosized materials might be related to the boundary conditions. The fact, we obtained $SiO_2$ nanoparticles by using two different boundary conditions, i.e. elastic reflection boundary conditions and the non-elastic reflection ones [15]. We found that structural properties of amorphous $SiO_2$ nanoparticles obtained by using two different boundary conditions are similar to each other and a significant discrepancy was found for the surface structure. Applying elastic reflection boundary conditions leads to an increase of structural defects at the surface of 2 nm nanoparticle compared with those observed in non-elastic ones. In contrast, no systematic changes have been found for 4 nm nanoparticle. In addition, we also found the boundary effects on the density profile of 4 nm nanoparticle. Such relatively small effects on static properties might lead to dramatic changes in the dynamics of atomic species in nanoparticles [15].

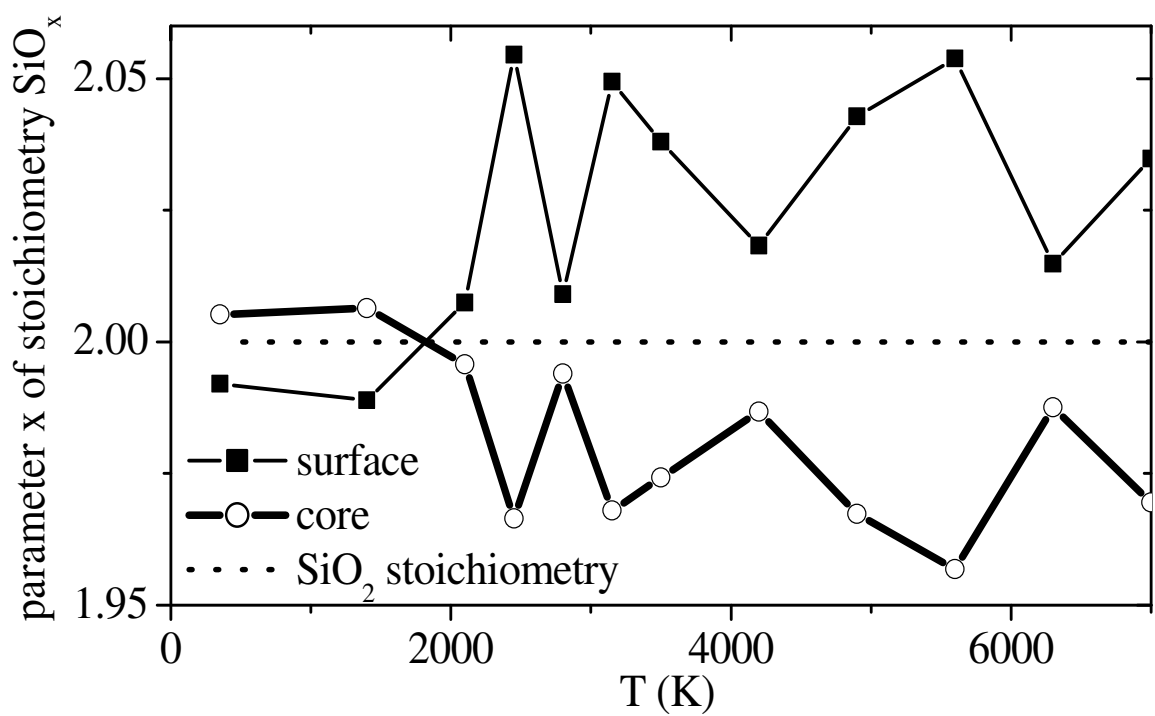

**Figure 6**
Temperature dependence of $SiO_x$ stoichiometry in the core and surface shell of $SiO_2$ nanoparticle with the size of 4 nm.

In addition, pressure-induced structural changes in nanoparticles are also of great interest. In particular, we found pressure-induced transition from low density form to high density one of amorphous $TiO_2$ nanoparticles [31]. Transition pressure of amorphous $TiO_2$ nanoparticles is size dependent in that the smaller nanoparticle size the higher the transition pressure like those commonly found in practice for nanocrystalline substances. Moreover, transition pressure of nanoparticles is higher than that of the bulk counterpart due to the surface effects [31]. On the other hand, so-called layer structure of amorphous $TiO_2$ nanoparticles is enhanced with increasing density of a model. The phenomenon oxygen atoms have a tendency to concentrate at the surface of amorphous nanosized substances has been found and it is strongly enhanced by pressurization in addition to the same but weaker tendency of Ti atoms. It causes a strong violation of local charge neutrality at the surface of nanoparticles, and consequently, in order to achieve the local charge neutrality Ti and O atoms also have a tendency to respectively concentrate in the inner shells. It leads to the forming so-called shell structure of compressed nanoparticles (i.e. respectively ordering spherical shells with Ti-rich and O-rich layers). Similar results for amorphous $SiO_2$ nanoparticles can be suggested since we found the same tendency caused by compression of initial 4 nm $SiO_2$ nanoparticle [32].

**Conclusion**

i) Calculations show that upon cooling from the melt toward amorphous state, tetrahedral network structure is forming in $SiO_2$ nanoparticles. Amorphous $SiO_2$ nanoparticles have a slightly distorted tetrahedral network structure if their size is large enough in that Si atoms are mainly surrounded by four O atoms like those observed in the bulk.

ii) The probability for the occurrence of structural defects in $SiO_2$ nanoparticles corresponding with $Z_{si-O} = 2$, $Z_{si-O} = 3$ and $Z_{O-Si} = 1$ have been described well by an Arrhenius law at high temperatures. However, prefactors $A_{ij}$ and activation energies $E_{ij}$ of the Arrhenius law for our models differ from those observed in the bulk.

iii) During the temperature decreasing, the increase of numerical density in the core of nanoparticle and its decrease in the surface layer are observed.

iv) Moreover, $SiO_x$ stoichiometry in the core and in the surface shell is also different from each other, i.e. while surface has a positive deviation from the $SiO_2$ stoichiometry, the core has a negative one over wide range of high temperature region (i.e. in the liquid state). The situation changes slightly at low temperatures. This means that in the high temperature region the core behaves oxygen-deficiency defects due to a negative deviation from the $SiO_2$ stoichiometry, while the surface shell has oxygen-excess defects due to a positive deviation from the $SiO_2$ one. This phenomenon may cause the appearance of additional defects in the $SiO_2$ nanoparticles.

v) In the present work, simulations were performed just at the ambient pressure density of 2.20 g/cm$^3$ for amorphous $SiO_2$, it is of interest to carry out the study at different densities ranged from low to high ones. Indeed, the pressure effects on static properties and dynamics of amorphous $SiO_2$ have been under intensive investigation [19,33]. In addition, it is better if one can carry out the quantum MD simulation for $SiO_2$ nanoparticles since the quantum effects become more important in the systems at nanoscale.

**Acknowledgements**
This work was accomplished in the Computational Physics Lab of the College of Natural Sciences – National University of HochiMinh City – Vietnam.